\documentclass[%
prl,twocolumn,longbibliography,
 amsmath,amssymb,
 aps,
]{revtex4-2}

\usepackage{bm}
\usepackage{graphicx}
\usepackage{enumitem}   
\usepackage{cleveref}
\graphicspath{{graphics/}}
\usepackage{float}
\usepackage{comment}
\usepackage{color}

\newcommand{\blue}{\textcolor{blue}}

\begin{document}

\widetext

\title{Optimizing the random search of a finite-lived target by a L\'evy flight}

\author{Denis Boyer$^1$}
\email{boyer@fisica.unam.mx}
\author{Gabriel Mercado-V\'asquez$^2$}
\email{gabrielmv.fisica@gmail.com}
\author{Satya N. Majumdar$^3$}
\email{satya.majumdar@universite-paris-saclay.fr}
\author{Gr\'egory Schehr$^4$}
\email{gregory.schehr@u-psud.fr}

\affiliation{$^1$Instituto de F\'\i sica, Universidad Nacional Aut\'onoma de M\'exico, Ciudad de M\'exico 04510, M\'exico}
\affiliation{$^2$Pritzker School of Molecular Engineering, University of Chicago, Chicago, IL, 60637, USA}
\affiliation{$^3$LPTMS, CNRS, Univ. Paris-Sud, Universit\'e Paris-Saclay, 91405 Orsay, France}
\affiliation{$^4$ Sorbonne Universit\'e, Laboratoire de Physique Th\'eorique et Hautes Energies, CNRS UMR 7589, 4 Place Jussieu, 75252 Paris
Cedex 05, France}

\date{\today}

\begin{abstract}
In many random search processes of interest in chemistry, biology or during rescue operations, an entity must find a specific target site before the latter becomes inactive, no longer available for reaction or lost. We present exact results on a minimal model system, a one-dimensional searcher performing a discrete time random walk or L\'evy flight. In contrast with the case of a permanent target, the capture probability and the conditional mean first passage time can be optimized. The optimal L\'evy index takes a non-trivial value, even in the long lifetime limit, and exhibits an abrupt transition as the initial distance to the target is varied.
Depending on the target lifetime, this transition is discontinuous or continuous, separated by a non-conventional tricritical point. 
These results pave the way to the optimization of search processes under time constraints.
\end{abstract}

\maketitle

Random search processes are ubiquitous in nature, such as animals searching for food \cite{bartumeus2009optimal,viswanathan2011physics}, rescue operations looking for survivors after a shipwreck \cite{serra2020search,kosmas2022saving} or even searches for a lost object like a key or a wallet. In typical search models, one considers the targets to be ``immortal'', {\it i.e.}, they do not disappear after a certain time. During the last decades, several models of random search of infinitely lived targets have been studied. The most popular among them is the search by a random walker, either diffusive or performing L\'evy flights where the jumps are long-ranged. Several strategies have been incorporated to make the search by a random walker optimal.
L\'evy walks with certain exponent values can maximize the capture rate by a forager of dispersed resources \cite{shlesinger1986levy,viswanathan1999optimizing,bartumeus2008fractal,levernier2020inverse,buldyrev2021comment,levernier2021reply,guinard2021intermittent,clementi2021search}.
Another well known strategy is the intermittent search process where short range and long range moves alternate to locate a single target \cite{oshanin2007intermittent,benichou2011intermittent}. A popular model that has received much attention in recent years
is a resetting random walker, where the walker goes back to its starting point with a finite probability 
after every step and restarts the search process~\cite{evans2011diffusion,evans2011diffusionb,reuveni2014role,roldan2016stochastic,pal2017first,falcon2017localization,evans2020stochastic,pal2022inspection}. In this case, it turns out that the mean time to find an infinitely lived target can be minimized by choosing an optimal resetting probability \cite{evans2011diffusion,evans2011diffusionb,evans2014diffusion,pal2016diffusion,reuveni2016optimal,montero2016directed,bhat2016stochastic,mercado2020intermittent,bressloff2020diffusive,de2022optimal,mercado2022reducing,biroli2023critical,evans2020stochastic}. This fact has also been verified in recent experiments in optical traps~\cite{tal2020experimental,besga2020optimal,faisant2021optimal}.

\begin{figure}
\centering			\includegraphics[width=.47\textwidth]{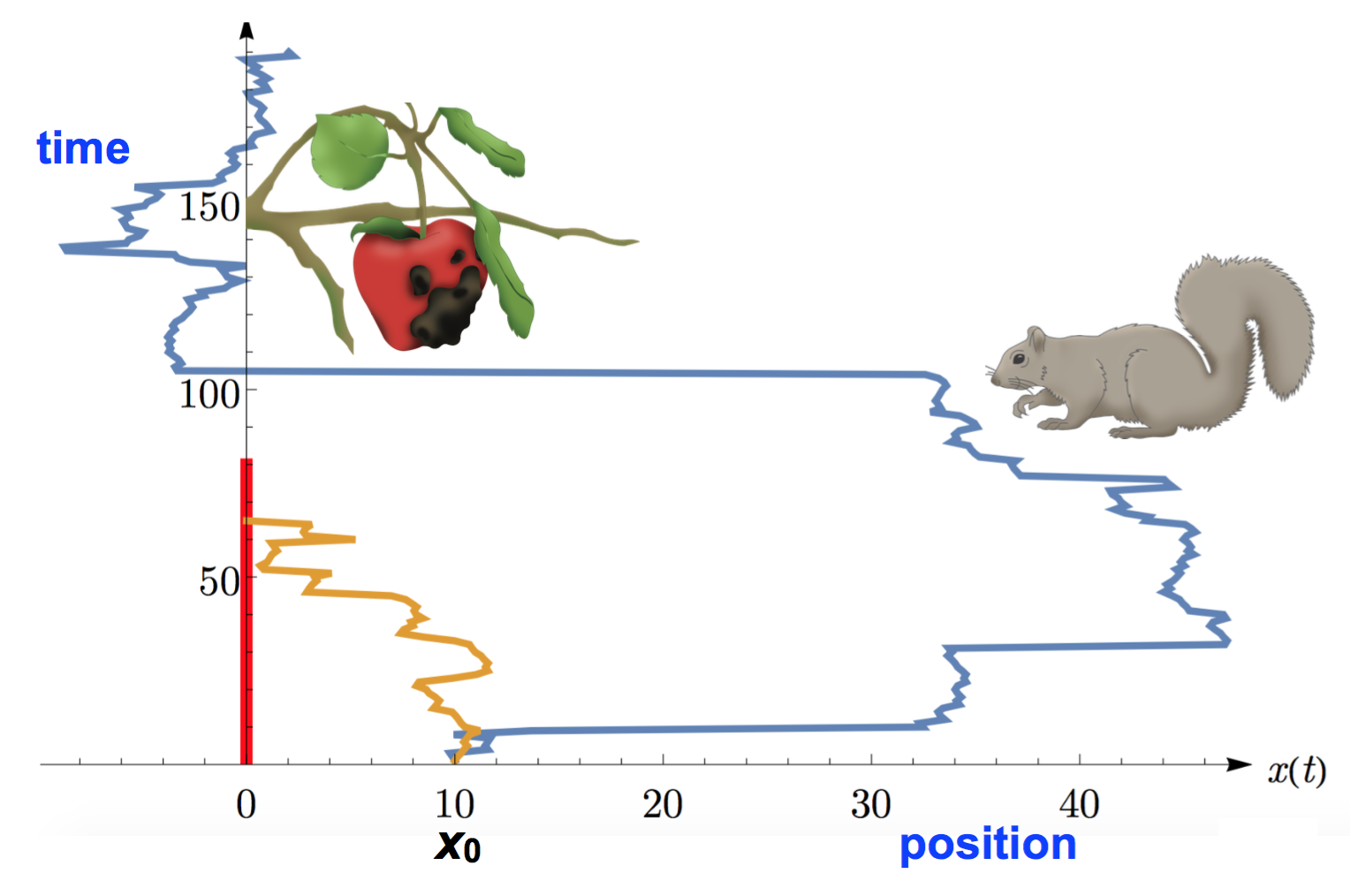}
		\caption{A searcher, performing a L\'evy flight in one-dimension, is looking for a non-permanent target ({\it i.e.}, a ripe fruit) located at the origin. At each time step, the target (in red) stays active with probability $a<1$, while the searcher performs a random step. If the searcher finds the target in the active state, the search is successful (orange trajectory). In contrast, if the target dies (rots) before being found by the searcher, the search is unsuccessful (blue trajectory).}
			\label{fig:cartoon}
\end{figure}

However, there are many instances where the target has a finite but random lifetime. For instance, ripe fruits in a tree rot in a few days. The lifetime of a fruit is typically random since it depends on the nature of the tree and the weather \cite{janmaat2006primates}. 
Similarly,  after a shipwreck, a survivor can last in the water only a finite amount of time, which is random as it depends on the general health of the person and sea conditions \cite{tikuisis1997prediction}. 
%
Inside a cell, target sites along the DNA are often blocked for long periods of time, which gives a limited random time to the transcription factors to bind to them~\cite{golding2005real,wong2008interconvertible,chen2014modulation}.
In many examples, the searcher has to capture the target before it disappears or dies. Alternatively, in a dual view, one can consider the target as permanent and the walker with a strong time constraint, as an aerial rescue vehicle having a limited flight time \cite{waharte2010supporting}. 
The termination of the search at a random time also appears in the context of foraging theory, where a searcher abandons a patch at any time with a certain give up probability~\cite{charnov1976optimal}.  
For a mortal searcher performing a lattice random walk \cite{yuste2013exploration} or Brownian motion \cite{meerson2015mortality}, the capture probability and conditional mean first-passage time cannot be optimized, or only with an infinite diffusion coefficient. If a resetting mechanism is further implemented, though, a non-zero resetting rate can be optimal provided the mortality rate is not too high \cite{belan2018restart,radice2023effects}.

A general question then is: is there any way to optimize the search success for a non-permanent target with a random lifetime? 
A natural generalization of the Brownian case is to investigate the search by a L\'evy flight with a L\'evy exponent $0 < \mu < 2$. One can then ask whether there is an optimal value of $\mu$ that minimizes the conditional search time or, alternatively, maximizes the capture probability of the mortal target. In this Letter, we address this problem for a one-dimensional L\'evy flight (see Fig. \ref{fig:cartoon}). In our model, the target is fixed at the origin and its lifetime $n$ is distributed geometrically via $p(n) = (1-a)\, a^n$ where $0<a<1$, i.e., at each discrete step, the target dies with probability $1-a$ and keeps alive with the complementary probability $a$. We assume that the L\'evy searcher starts from $x_0>0$ and subsequently evolves in discrete time via
\begin{equation}\label{walk}
    x_n=x_{n-1}+\eta_n 
\end{equation}
where $\eta_n$'s are independent and identically distributed  jump variables, each distributed via the probability distribution function  
$f(\eta)$, which we assume to be symmetric and continuous with a power-law tail $\propto 1/|\eta|^{1+\mu}$ where $\mu\in(0,2)$.
Note that both parameters $x_0$ and $a$ are given numbers and the searcher has no control in optimizing with respect to them. Thus the only parameter that the searcher has in her disposal to optimize is $\mu$, since it is associated with her motion. The search is successful only if the walker crosses the origin for the first time (takes $x_n<0$) while the target is still alive. We characterize the search success by two different observables: {\it (i)} the capture probability of the target and {\it (ii)} the conditional mean first-passage time (CMFPT), {\it i.e.}, the mean search time conditioned to finding the target alive. We find that, for fixed $x_0$ and $a$, these two quantities can be optimized by varying the L\'evy index $\mu$. 
The two optimal parameters $\mu_{cap}^{\star}(x_0,a)$ and $\mu_{FP}^{\star}(x_0,a)$ exhibit very rich phase diagrams in the $(x_0,a)$ plane. 

\begin{figure}
\centering			\includegraphics[width=.49\textwidth]{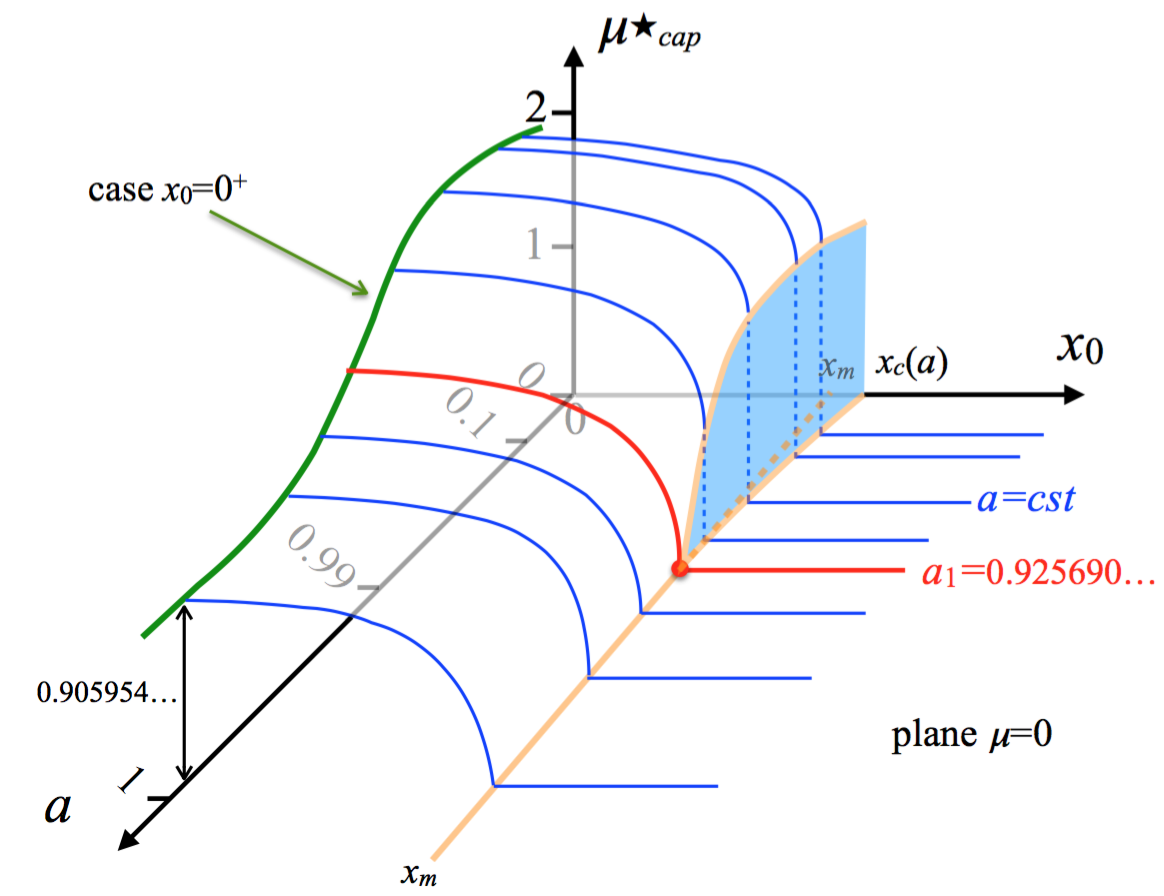}
		\caption{Schematic phase diagram of the optimal L\'evy index $\mu_{cap}^*$ in the $(x_0,a)$ plane. For fixed $a$, as a function of $x_0$, the optimal $\mu_{cap}^*$ undergoes a first-order transition at $x_0=x_c(a)$ (for $a<a_1$) which changes to a $2^{\rm nd}$-order transition for $a>a_1$. The critical line $x_c(a)$ freezes to $x_m = 0.561459\ldots$ for $a>a_1$. The point that separates the first-order and second-order transitions is a tricritical point (shown by the red dot).}
			\label{fig:diag}
\end{figure}

Our results, obtained analytically and numerically, are summarized schematically in Fig.~\ref{fig:diag} for the capture probability. 
For any fixed $a<a_1= 2\,e\,(\sqrt{15}-2)/11 = 0.925690\ldots$, the index
$\mu^{\star}_{cap}(x_0,a)$ decreases monotonically as a function of $x_0$, and jumps to zero abruptly at a critical value $x_0=x_c(a)$. This signals a first-order transition. In contrast, for any $a>a_1$, $\mu^{\star}_{cap}(x_0,a)$ again decreases with $x_0$ but vanishes continuously at $x_c(a)$, signaling a second-order transition. 
In the case $a > a_1$, the critical value  $x_c(a)$ freezes to a constant value $x_c(a) = x_m=0.561459\ldots$.
Thus $(x_m,a_1)$, shown by a red dot in Fig. \ref{fig:diag}, is a tricritical point that sits at the junction of a $1^{\rm st}$ and $2^{\rm nd}$-order transition. The green line $x_0\to0$ is obtained analytically in the Supplemental Material \cite{SM}. A qualitatively similar diagram can be drawn for $\mu^{\star}_{FP}(x_0,a)$, with 
a tricritical point at a slightly larger value $a_2=0.973989\ldots$ \cite{SM}.

\begin{figure*}[t]
\includegraphics[width=.32\textwidth]
{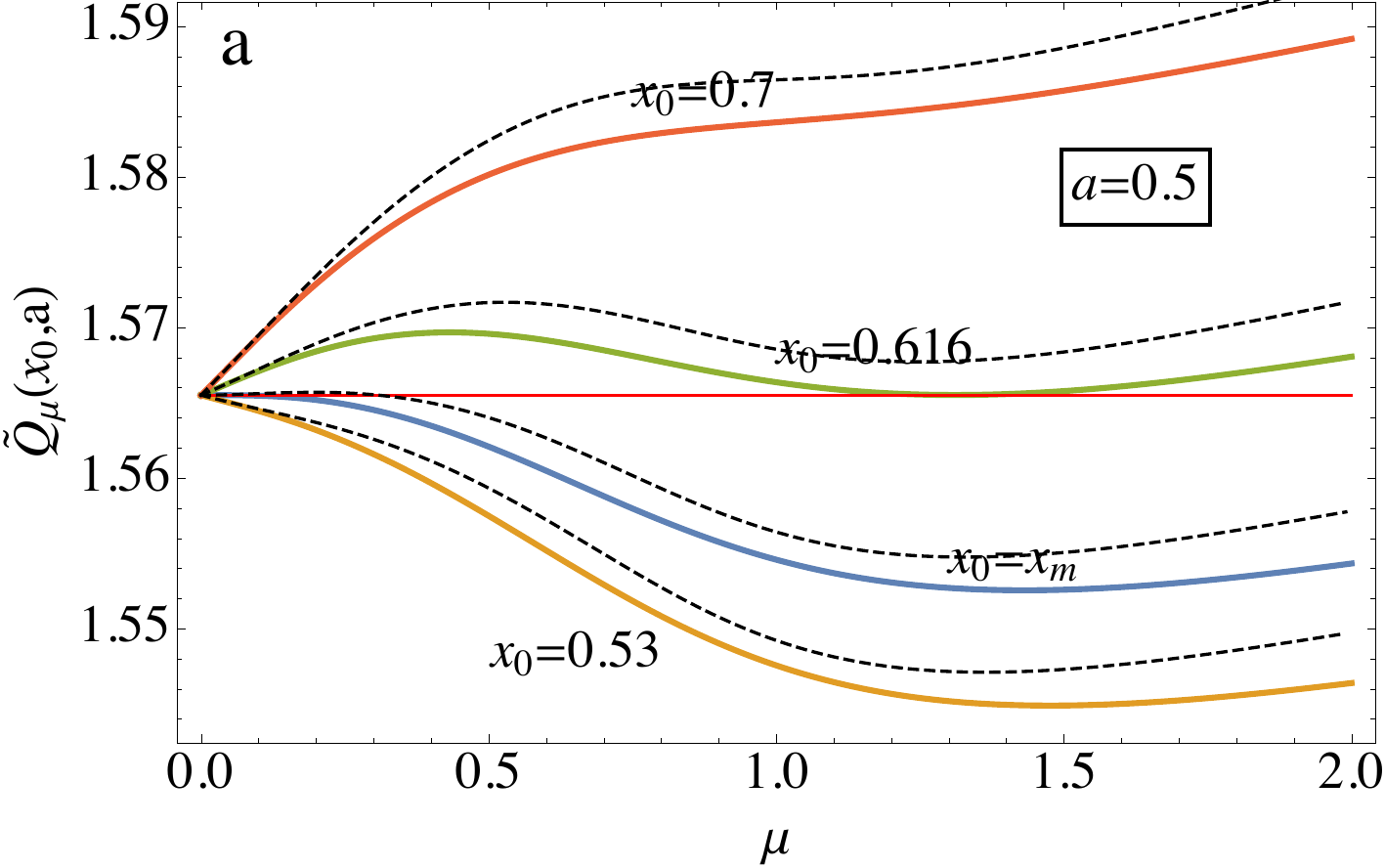}
\includegraphics[width=.32\textwidth]
{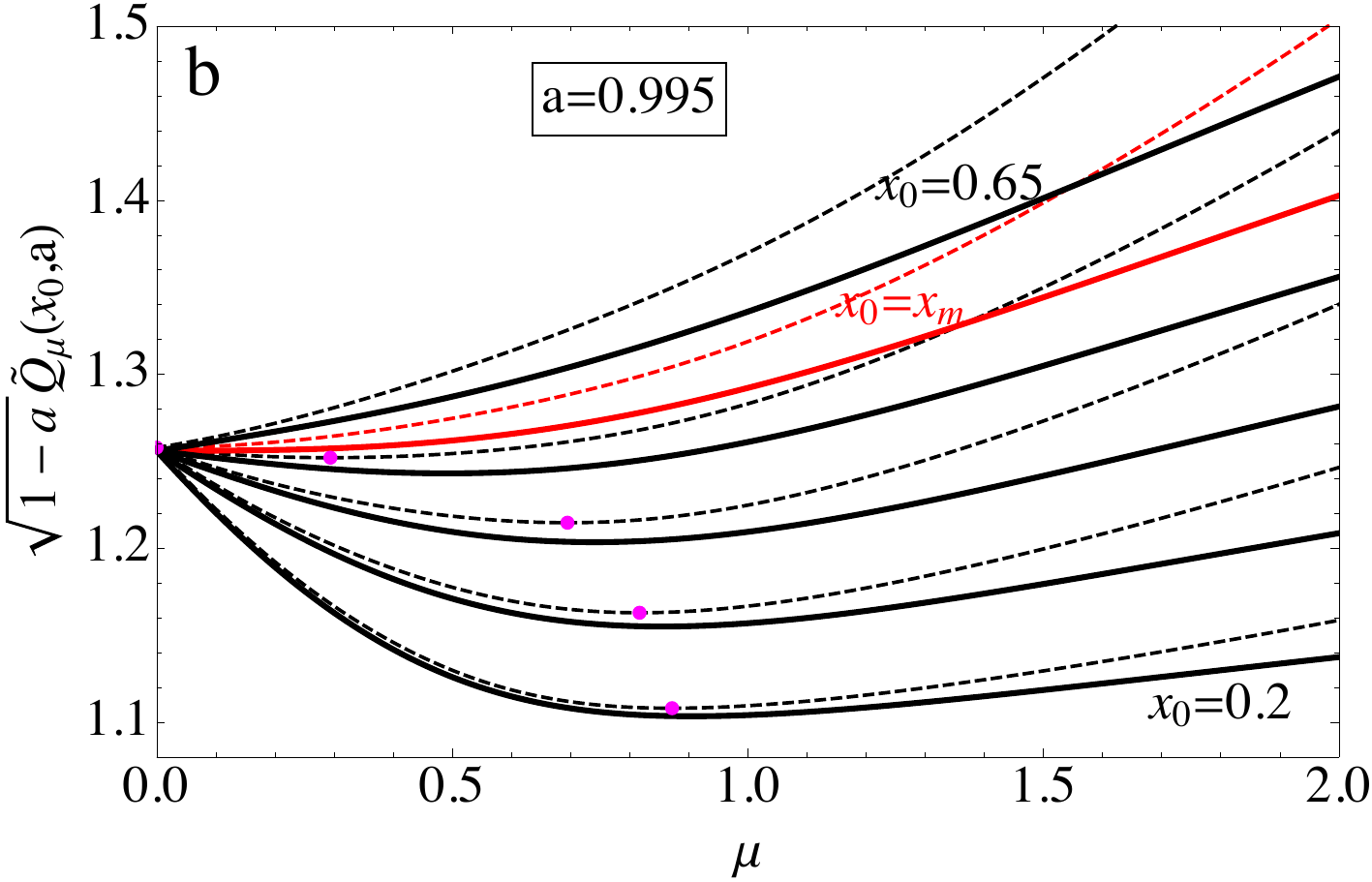}
\includegraphics[width=.32\textwidth]
{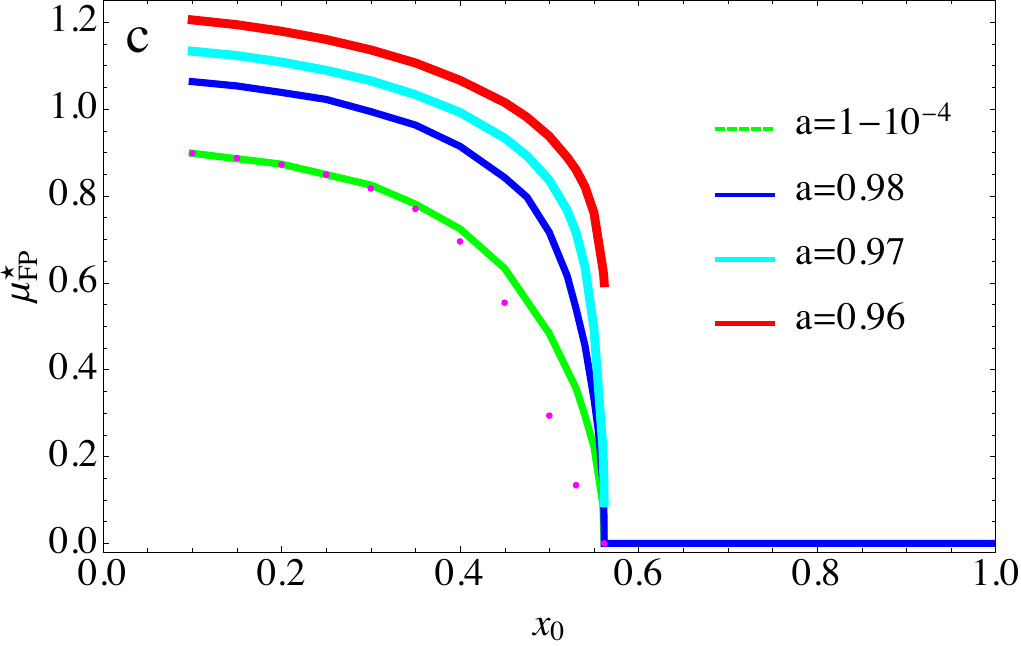}
		\caption{(a) Discontinuous transition with short-lived targets ($a=0.5$): numerical $\widetilde{Q}_{\mu}(x_0,a)$ vs. $\mu$ for different starting positions close to $x_m$. (b) Continuous transition for long-lived targets ($a$ close to 1): $\sqrt{1-a}\widetilde{Q}_{\mu}(x_0,a)$ as a function of $\mu$ and for several $x_0$ around $x_m$.  In (a) and (b) the dotted lines represent the concavity approximation (\ref{concav}).
  (c) Optimal exponent for the CMFPT as a function of $x_0$ for various $a$. Below $a_2=0.973989...$  the transition is discontinuous ($a=0.97$), while it is continuous above ($a=0.98$). The dots correspond to the minima in (b). The index $\mu^{\star}_{cap}(x_0,a)$ has analogous variations near $a_1$.}
		\label{fig:Ta1}
\end{figure*}

Both observables, the capture probability and the CMFPT, can be related to one fundamental quantity $Q_{\mu}(x_0,n)$ 
associated with the random walk, denoting the probability that a L\'evy walker with index $\mu$, starting at $x_0\geq 0$, does not cross $0$ up to step $n$~\cite{pollaczek1952functions,andersen1954fluctuations,spitzer1956combinatorial,redner2001guide,MC_precise,SNM:Leuven,BMS_review,Majumdar_2017,kusmierz2014first,klinger2022splitting}. Consequently, $Q_\mu(x_0,n-1) - Q_\mu(x_0,n)$ is the probability that the L\'evy flight crosses the origin for the first time at the $n$-th step, with $Q_{\mu}(x_0,n=0)=1$. Thus for the target to be captured at the $n$-th step, it has to remain alive at least up to step $n-1$, which occurs with probability $a^{n-1}$. Therefore the capture probability $C_{\mu}(x_0,a)$, defined as the probability that the searcher starting at $x_0$ finds the target before the latter becomes inactive, is given by $C_{\mu}(x_0,a)=\sum_{n=1}^{\infty} a^{n-1}\left[Q_\mu(x_0,n-1)-Q_{\mu}(x_0,n)\right]$. This sum can be rewritten as
\begin{equation}\label{Cgen}
C_{\mu}(x_0,a)=\frac{1-(1-a)\widetilde{Q}_\mu(x_0,s=a)}{a},
\end{equation}
where $\widetilde{Q}_\mu(x_0,s)\equiv\sum_{n=0}^{\infty} s^nQ_{\mu}(x_0,n)$ is the generating function of $Q_{\mu}(x_0,n)$. 
Similarly, the CMFPT $T_\mu(x_0,a)$, the mean time taken by the successful trajectories to locate the target~\cite{meerson2015mortality}, can be expressed as $T_\mu(x_0,a)=\sum_{n=1}^{\infty} n a^{n-1}\left[Q_\mu(x_0,n-1)-Q_{\mu}(x_0,n)\right]/C_{\mu}(x_0,a)$, where $C_{\mu}(x_0,a)$ acts as a normalization factor. This can also be rewritten again in terms of the generating function of the survival probability 
\begin{equation}
T_\mu(x_0,a)
=a\frac{\partial}{\partial a}\ln\left[1-(1-a)\widetilde{Q}_\mu(x_0,s=a) \right].\label{Tgen2}
\end{equation}
%

Thus to analyze either $C_\mu(x_0,a)$ or $T_\mu(x_0,a)$, we need the generating function $\widetilde Q_\mu(x_0,s)$ for L\'evy flights. Unfortunately, there is no simple expression for $\widetilde Q_\mu(x_0,s)$. However its Laplace transform with respect to $x_0$ is given by the exact Pollaczek-Spitzer formula~\cite{pollaczek1952functions,spitzer1956combinatorial},
\begin{eqnarray}\label{PS}
&&\int_0^{\infty}\widetilde{Q}_\mu(x_0,s) \, e^{-\lambda x_0} dx_0=
\frac{1}{\lambda\sqrt{1-s}} \varphi(\lambda,s) \\
&& {\rm with}\quad \varphi(\lambda,s)  = \exp\left[ -\frac{\lambda}{\pi}
\int_0^{\infty}\frac{\ln[1-s\hat{f}(k)]}{\lambda^2+k^2} dk\right] \,, \quad
\end{eqnarray}
where $\hat{f}(k)=\int_{-\infty}^{\infty}f(\eta)\ e^{{\rm i}k\eta}d\eta$ is the Fourier transform of the step distribution. Here we will focus on L\'evy stable jump distribution, with $\hat{f}(k)=e^{-|k|^{\mu}}$ with $0<\mu\le 2$.

With an infinite-lived target ($a=1$), recall that $C_{\mu}=1$, owing to the recurrence property of $1d$ random walks, while $T_{\mu}=\infty$, independently of $x_0$ and $f(\eta)$~\cite{Fellerv1}. Hence, there is no option of optimizing them by varying $\mu$. However, for a finite-lived target where $a<1$, both quantities become nontrivial functions of $\mu$ and can be optimized by choosing $\mu$ appropriately with optimal values 
$\mu^{\star}_{cap}(x_0,a)$ and $\mu^{\star}_{FP}(x_0,a)$. One finds that, even for short-lived targets, $C_{\mu}$ at optimality can be larger than the maximal value $1/2$ that could be achieved by a naive ballistic strategy (see \cite{SM}).

In order to maximize the capture probability in Eq.~\eqref{Cgen} by varying $\mu$, for fixed $x_0$ and $a$, it turns out that we need to minimize  $\widetilde Q_{\mu}(x_0,s=a)$ with respect to $\mu$. We will study the exact relation in Eq. \eqref{PS}, both analytically in certain limits and numerically by inverting the Laplace transform in Eq. \eqref{PS} 
using the Gaver-Stehfest method \cite{StehfestNumLaplaceInv,kuhlman2013review}, which we explain in \cite{SM}.

We start by plotting the numerically obtained  $\widetilde{Q}_\mu(x_0,a)$ as a function of $\mu$, for fixed $x_0$ and $a$. In Fig. \ref{fig:Ta1}a we show the data for $a=0.5$ and four different values of $x_0$. For small $x_0$, the curve has a single minimum at a nonzero value of $\mu^{\star}_{cap}(x_0,a)$, while there is a local maximum at $\mu=0$. As $x_0$ increases to some value $x_m$, the derivative of $\widetilde{Q}_\mu(x_0,a)$ with respect to $\mu$ at $\mu = 0^+$ \footnote{Here we consider the $\mu \to 0^+$ limit (and not strictly $\mu=0$). In the limit $\mu \to 0^+$ the jump distribution is normalizable but not when $\mu=0$ exactly. Hence we restrict only to the case $\mu \to 0^+$.} vanishes, {\it i.e.}, 
$\partial_\mu \widetilde Q_\mu(x_m,a) \Big \vert_{\mu = 0} = 0 \;.$
This value of $x_m$ can be determined analytically [see Eq. (\ref{q1}) below] and is given by $x_m = e^{-\gamma_E} = 0.561459\ldots$, where $\gamma_E$ is the Euler constant. 
When $x_0$ slightly exceeds $x_m$, the curve has two minima: one at $\mu=0^+$ and one at $\mu = \mu^{\star}_{cap}(x_0,a)$, but the value at $\mu=0^+$ is higher. This situation persists for $x_m < x_0 < x_c(a)$. When $x_0$ exceeds $x_c(a)$, the local minimum at $\mu=0^+$ becomes the global one and $\mu^{\star}_{cap}(x_0,a)$ drops to $0^+$, triggering a first-order transition.  The point $x_c(a)$ is thus determined by 
\begin{equation} \label{def_xc}
\left.\partial_{\mu}\widetilde{Q}_{\mu}(x_c,a)\right|_{\mu^{\star}_{cap}(x_c)}=0,\quad  
\left.\widetilde{Q}_{\mu}(x_c,a)\right|_{\mu^{\star}_{cap}(x_c)}=q_0,
\end{equation}
where $q_0\equiv\widetilde{Q}_{\mu=0}(x_c,a)$. From Eq. (4), $q_0=1/\sqrt{(1-a)(1-ae^{-1})}$, independent of the position (see \cite{SM}).
This scenario presented above for $a=0.5$ continues to hold up to $a=a_1 \approx 0.926$. 

For $a>a_1$, a different scenario occurs as displayed in Fig.~\ref{fig:Ta1}b where again $\widetilde Q_\mu(x_0,a)$ is plotted as a function of $\mu$ for different values of $x_0$. In contrast to Fig. \ref{fig:Ta1}a, the curves always have a single minimum at $\mu = \mu^{\star}_{cap}(x_0,a)$ that decreases continuously to \blue{$0^+$} as $x_0$ approaches a critical value $x_c(a)=x_m$, signaling a second-order phase transition. Thus the first and second-order phase transitions merge at $a=a_1$, making it a tricritical point.  These numerical observations lead to the phase diagram presented in Fig. \ref{fig:diag}. 

The CMFPT exhibits the same qualitative features as above, with a tricritical point now located at $a=a_2\approx 0.974...$ In 
Fig.~\ref{fig:Ta1}c, we plot $\mu^{\star}_{FP}(x_0,a)$ as a function of $x_0$ for four different values of $a$ close to $a_2$. The jump discontinuity at $x_0=x_c(a)$ is finite for $a<a_2$ while it vanishes for $a\geq a_2$, confirming indeed that $(x_0=x_m,a=a_2)$ is a tricritical point for $\mu^{\star}_{FP}(x_0,a)$ in the $(x_0,a)$ plane.

We show how $a_1$ and $a_2$ can be computed analytically using a standard Landau-like expansion well known in critical phenomena. There, by expanding the free energy in powers of the order parameter, the Landau theory gives access to the phase diagram close to a continuous critical/tricritical point. Here we follow the same spirit with $\mu$ playing the role of the ``order parameter''. We then expand $\widetilde{Q}_{\mu}$ in powers of $\mu$ near $\mu=0^+$: $\widetilde{Q}_{\mu}(x_0,a)=q_0+q_1\mu+q_2\mu^2/2!+q_3\mu^3/3!+q_4 \mu^4/4! + \ldots$, where the dependence of the $q_i$'s on $x_0$ and $a$ is implicit. Depending on these parameters, the signs of $q_i$'s in this expansion may change, leading either to a first or second order transition and also to the possibility of a tricritical point. In the standard Landau's theory with a positive order parameter it is enough to keep terms up to order $O(\mu^3)$ and a tricritical point occurs when $q_1 = q_2 = 0$ while $q_3>0$ \cite{kincaid1975phase} (see also~\cite{pal2019landau} in the context of stochastic resetting). However, in our case, the dependence of $q_i$'s on $x_0$ and $a$ are such that this standard scenario is not realized and one needs to keep terms up to order $O(\mu^4)$. From Eq. (\ref{PS}), we show that \cite{SM}
\begin{eqnarray}
&&q_1=\frac{ae^{-1}}{2\sqrt{1-a}(1-ae^{-1})^{3/2}}(\ln x_0+\gamma_E) \label {q1} \\
&&q_2=\frac{3\sqrt{e}a^2}{4\sqrt{1-a}(e-a)^{5/2}}(\ln x_0+\gamma_E)^2 \label{q2} \;.
\end{eqnarray}
For $x_0<x_m = e^{-\gamma_E}$, we have $q_1 < 0$ and $q_2>0$. In contrast, for $x_0>x_m$, we have both $q_1, q_2 > 0$ and both of them vanish simultaneously at $x_0=x_m$, for any $a$. The tricritical point thus occurs when $q_3(x_m,a)$ changes sign. We have \cite{SM}
\begin{equation}\label{q3}
q_3(x_m,a)=\frac{a\sqrt{e}K}{8\sqrt{1-a}(e-a)^{7/2}}
(11a^2+8ea-4e^2) \;,
\end{equation}
where $K=2\zeta(3) = 2.40411...$. Thus $q_3(x_m,a) < 0$ for $a<a_1$ where $a_1=2e(\sqrt{15}-2)/11$ is the unique root of $11a^2+8ea-4e^2=0$ in $(0,1)$. 
At the transition point $x_0=x_c(a)$ and for $a<a_1$, since $q_3<0$, we need to keep terms up to order $O(\mu^4)$ (assuming that $q_4>0$ in the Landau expansion). From Eqs. (\ref{def_xc}), the first-order jump discontinuity 
 $\Delta(a)\equiv\mu_{cap}^{\star}(x_c(a),a)$ is given by \cite{SM}
\begin{eqnarray}\label{Delta_text}
\Delta(a) = \frac{2}{3 \, q_4} \left( 2|q_3| + \sqrt{4q_3^2 - 9 q_2 q_4} \right)\Big \vert_{x_0=x_c(a)} \;,
\end{eqnarray} 
This discontinuity vanishes when $q_3 \to 0$ and $q_2 \to 0$ which occurs at the point $(x_0 =x_m, a=a_1)$, indicating that this is a tricritical point. If $a>a_1$ then $q_2>0$ and $q_3>0$ : when $q_1$ changes sign (always at $x_0=x_m$), a $2^{\rm nd}$ order transition occurs. Hence $x_c(a)$ freezes to $x_m$ for $a>a_1$.  
A similar Landau-like expansion can be carried out exactly for the CMFPT, which leads to the same qualitative conclusions, with $a_2 = 0.973989\ldots$ \cite{SM}.

As mentioned before, for a permanent target ($a=1$), there is no optimal strategy since the capture probability is $1$ and the CMFPT infinite, irrespective of $\mu$. However, surprisingly, for long-lived targets, there is a nontrivial optimal strategy characterized by the same
$\mu_{cap}^* = \mu_{FP}^*$ for both observables. 
As $a \to 1$, Eqs. (\ref{PS}) and (\ref{Tgen2}) imply $\widetilde Q_\mu(x_0,a) \approx g_\mu(x_0)/\sqrt{1-a}$ and $T_{\mu(x_0,a)}\approx g_\mu(x_0)/(2\sqrt{1-a}$), where $g_\mu(x_0)$ is independent of $a$. Hence, both the capture probability and the CMFPT are optimized by minimizing $g_\mu(x_0)$ with respect to $\mu$. Since the expression of $g_\mu(x_0)$ is complicated, it is hard to obtain the full functional form of $\mu_{cap}^* = \mu_{FP}^*$ for all $x_0$. However, close to the transition point $x_m$, where $\mu_{cap}^*$ is expected to be small due to the continuous transition, $g_{\mu}$ directly follows from the small $\mu$ expansion of $Q_\mu$ above. Using Eqs. (\ref{q1}) and (\ref{q3}), we obtain exactly to leading order for small $(x_m-x_0)/x_m$
\begin{eqnarray}\label{small_eps}
\mu_{cap}^* =\mu_{FP}^{\star}\approx A \left( \frac{x_m-x_0}{x_m}\right)^{1/2} \;, \; x_0 < x_m \;,
\end{eqnarray}  
where $A = 2(e-1)/\sqrt{\zeta(3)(11+8e-4e^2)} = 1.7549\ldots$ (see SM \cite{SM} for more details). This shows that the limit $a \to 1$ does allow an optimization with respect to $\mu$.

So far, we have analyzed the exact formula in Eq.~\eqref{PS} in the small $\mu$ limit. 
When $a\to1$ and $x_0\to0$, far from $x_m$, a small $x_0$ expansion in \cite{SM} gives $\mu^{\star}_{cap}\to 0.905954...$, as indicated in Fig. \ref{fig:diag}. But
to obtain analytically the full curves in Figs. \ref{fig:Ta1}a and \ref{fig:Ta1}b, as a function of $\mu$ from Eq. \eqref{PS} for any $(x_0,a)$ is extremely hard. Yet, we have found a concavity approximation allowing a very accurate analytical estimate of $\widetilde Q_\mu(x_0,a)$.  
Starting from the concavity of the logarithm, we approximate  $\sum_i w_i \ln(r_i) \approx \ln( \sum_i w_i r_i)$ for any set of positive reals $r_i$ and normalized weights  $\sum_i w_i = 1$. With this, one can perform the inverse Laplace transform in Eq. (\ref{PS}) and deduce the general expression
\begin{equation}\label{concav}
\widetilde{Q}_{\mu,approx}(x_0,s)= \frac{1}{\sqrt{1-s}}e^{-\frac{1}{\pi}\int_0^{\infty}\ln[1-s\hat{f}(k)]\frac{\sin(kx_0)}{k}dk} \;,
\end{equation}
where we have used the identity ${\cal L}^{-1}[k/(\lambda^2+k^2)]=\sin(kx_0)$ for $x_0>0$ (see also \cite{SM}). Eq. (\ref{concav}) is easy to evaluate numerically. Interestingly, the first two terms of its small $\mu$ expansion coincide with the exact expressions $q_0$ and $q_1$ above, as well as the first terms of its small $x_0$ expansion \cite{SM}. Consequently, Eq. (\ref{concav}) gives the correct slope-change point $x_m$ and also captures qualitatively the order of the transitions (see the dashed lines in Figs. \ref{fig:Ta1}a and \ref{fig:Ta1}b), along with the existence of a tricritical point.

We conclude with the remark that this problem of a finite-lived target is reminiscent of a L\'evy flight subject to resetting with a probability $r$ to its initial position. The mean first-passage time (MFPT) to find a permanent target at the origin was computed for the resetting L\'evy flight~\cite{kusmierz2014first} where the walker has two parameters $\mu$ and $r$ that can be used to optimize the MFPT (see also \cite{campos2015phase} for a related problem). Indeed, the optimal pair $(\mu^*,r^*)$ was computed and found to undergo a first-order transition at a critical value of the initial distance $x_0$ from the target. This is rather different from our problem where
the L\'evy flight has only a single parameter $\mu$, which it can vary to optimize the MFPT. In our model, the walker has no control over the parameter $a$ associated with the lifetime of the target. Hence, here we optimize the search strategy by varying only $\mu$ for {\it fixed} $a$, which leads to a new phase diagram with a tricritical point.

In summary, we have studied a simple model of a L\'evy flight of index $\mu$ in one-dimension searching for a {\it finite-lived} target at the origin with mean lifetime $1/(1-a)$. We have shown that the capture probability of the target can be maximized by choosing an optimal $\mu^{\star}_{cap}$ for fixed $a$ and $x_0$ (where $x_0$ denotes the initial distance from the target). The presence of a finite life-time leads to a very rich and nontrivial phase diagram for $\mu^{\star}_{cap}$ in the $(x_0,a)$ plane. This work opens up many interesting possibilities for future works. For instance, it would be interesting to find the optimal strategy in higher dimensions, for multiple L\'evy flights and for the case where the distribution of the target lifetime is non-exponential. \\

DB acknowledges support from the LPTMS at the Universit\'e Paris-Saclay (France) and from CONACYT (Mexico) grant Ciencia de Frontera 2019/10872. We thank Lya Naranjo for illustration support in Fig. 1.


%
\end{document}


\begin{center}
{\Large Supplemental Material of\\
\lq\lq Optimizing the random search of a finite-lived target by a L\'evy flight"}\\
 D. Boyer, G. Mercado-V\'asquez, S. N. Majumdar and G. Schehr
\end{center}

\vspace{1cm}

\section{Numerical inversion of Laplace transforms}\label{appStehfest}

The generating function $\widetilde{Q}_{\mu}(x_0,s)$ appearing in the Pollaczek-Spitzer formula given by Eqs. (4)-(5) of the Main text was numerically computed by means of the Gaver-Stehfest method. This methods provides a rapid and easy-to-implement way of inverting Laplace transforms \cite{StehfestNumLaplaceInv,kuhlman2013review}. An arbitrary function $G(t)$ is approximately deduced from its Laplace transform $g(s)=\int_{0}^\infty e^{-st}G(t)dt$ as \cite{kuhlman2013review}
\begin{equation}
    G(t)=\frac{\ln 2}{t}\sum_{k=1}^N V_k\ g\left(k\frac{\ln 2}{t}\right)\label{Stehfest}
\end{equation}
where 
\begin{equation}
    V_k=(-1)^{k+N/2}\sum_{j=\lfloor (k+1)/2\rfloor}^{\mathrm{min}(k,N/2)}\frac{j^{N/2}(2j)!}{(N/2-j)!j!(j-1)!(k-j)!(2j-k)!}.
\end{equation}
The number $N$ above must be even and its choice depends on the behavior of $g(s)$. Generally, it is sufficient to set $N\le 16$ for achieving double precision. After testing several values of $N$ we found that $N=6$ was good enough to compute a solution with the desired accuracy in our problem. 

We recall that the Pollaczek-Spitzer formula is expressed as
\begin{equation}\label{PS2}
\int_0^{\infty}\widetilde{Q}_{\mu}(x_0,s)e^{-\lambda x_0} dx_0=
\frac{1}{\lambda\sqrt{1-s}}\exp\left[ -\frac{\lambda}{\pi}
\int_0^{\infty}\frac{\ln[1-s\hat{f}(k)]}{\lambda^2+k^2}dk\right],
\end{equation}
where $x_0>0$ is the starting position of the searcher.
Prior to the inversion as in Eq. (\ref{Stehfest}), one needs to compute the Fourier integral in the right side of Eq. (\ref{PS2}), assuming that the Fourier transform $\hat{f}(k)$ of the step distribution is known. To avoid dealing with the infinite integration domain and to reduce the accumulation of errors, for the L\'evy stable distribution with $\hat{f}(k)=e^{-k^{\mu}}$ we can make the change of variable $u=e^{-k^{\mu}}$ and express the integral as
\begin{equation}
    \int_0^{\infty}\frac{\ln[1-s\hat{f}(k)]}{\lambda^2+k^2} dk=\int_0^{1}\frac{\ln[1-su]}{\mu\left[\lambda^2+\left(-\ln u\right)^{2/\mu}\right]\left(-\ln u\right)^{\frac{\mu-1}{\mu}}u} du,\label{intApp}
\end{equation}
which is well defined in the interval $\mu\in(0,2]$. The singularity at $\mu=0$ is computed analytically and inverted as $q_0(x_0,s)$ in Eq. (\ref{q0}) below.
The integral (\ref{intApp}) can now be solved by standard algorithms or with Mathematica. For each different value of $x_0$ or $\mu$ in which the Pollaczek-Spitzer formula is to be inverted, $N$ integrals must be calculated. The Gaver-Stehfest method is not too time-consuming and remains suitable for optimization with respect to $\mu$.

 \section{Landau-like expansion of $\widetilde{Q}_{\mu}(x_0,s=a)$ in powers of $\mu$}\label{LandauQ}

Let us set $s=a$ in Eq. \eqref{PS2}, where $a$ is the living probability of the target, and consider the expansion of $\widetilde{Q}_{\mu}(x_0,a)$ in powers of $\mu$ near $\mu=0$,
\begin{equation}\label{Qq}
\widetilde{Q}_{\mu}(x_0,a)=q_0(x_0,a) +q_1(x_0,a)\mu+\frac{1}{2!}q_2(x_0,a)\mu^2+\frac{1}{3!}q_3(x_0,a)\mu^3+\frac{1}{4!}q_4(x_0,a)\mu^4 + \cdots \;.
\end{equation}
In the subsection following the next one, we will calculate the coefficients $q_i$'s explicitly.  In the next subsection, we consider the general scenario, assuming that the $q_i$'s are given. 

\subsection{General consideration}

In the standard Landau's theory of tricritical point with  positive order parameter, it is usually enough to keep terms only to order $O(\mu^3)$, but here we will see that we need to keep terms up to order $O(\mu^4)$. 
\begin{figure}[t]
\includegraphics[width = 0.8\linewidth]{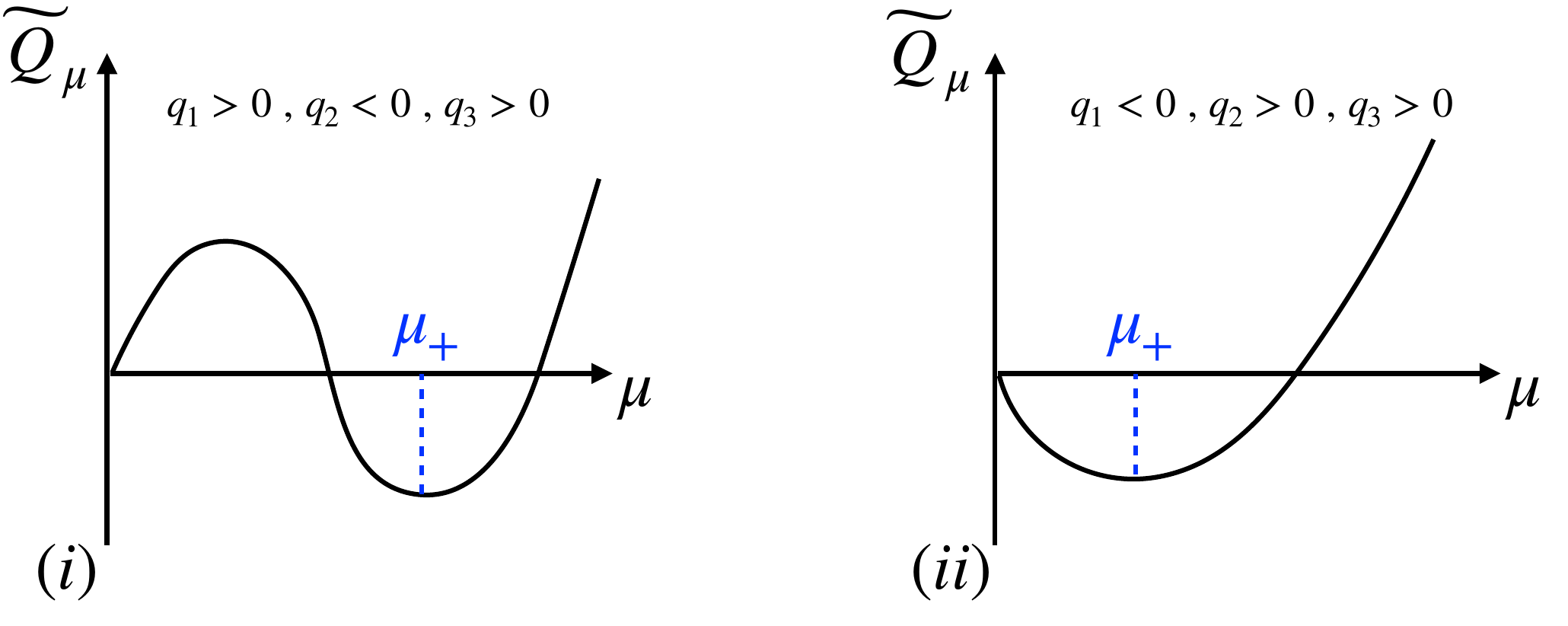}
\caption{{\bf Left panel:} A schematic plot of $\widetilde Q_\mu$ vs $\mu$ in Eq. \eqref{Qq} with $q_0=0$ and keeping terms up to order $O(\mu^3)$ for the case when $q_1>0, \; q_2<0$ and $q_3>0$. {\bf Right panel:} the same curve for $q_1<0, \; q_2 >0$ and $q_3>0$.}\label{Fig_tri}
\end{figure}
It is useful to first recall briefly the standard theory of tricritical phase diagram within the Landau expansion, where $\mu$ is treated as the order parameter and one restricts the expansion up to order $O(\mu^3)$. 
Usually, there are two competing scenarios leading to either a $(i)$ first-order or $(ii)$ a second-order transition, depending on the coefficients $q_1, q_2$ and  $q_3$. We consider the two scenarios separately:
\begin{itemize}
\item[$(i)$]{Consider the case when $q_1>0$ and $q_3>0$ and $q_2<0$. In this case,  the curve $\widetilde{Q}_\mu$ as a function of $\mu$ is schematically shown in the left panel of Fig. \ref{Fig_tri}. Here the function has a global minimum at $\mu=\mu_+$. As we change the coefficients (for instance the slope at the origin), the value of the global minimum increases and, at some point, it hits the value $0$ where the minimum at $\mu=\mu_+$ and the one at $\mu=0$ coexist. Beyond this point, the minimum at $\mu=0$ is the global minimum. Thus the value of the order parameter jumps from $\mu_+$ to $0$, signalling a first-order phase transition. Exactly at the transition point (when the two minima coexist), we thus have two conditions 
\begin{eqnarray}\label{cond_first}
\widetilde{Q}_{\mu=\mu_+} = \widetilde Q_{\mu = 0} \; &\Longrightarrow& \; q_1 \mu + q_2 \frac{\mu^2}{2} + q_3 \frac{\mu^3}{6} = 0 \\
\frac{\partial \widetilde Q_{\mu}}{\partial \mu} \Big \vert_{\mu =0} = 0 \; &\Longrightarrow &\; q_1 + q_2 \mu + q_3 \frac{\mu^2}{2} = 0 \;.
\end{eqnarray}	 
Solving this pair of equations, one gets the jump at the first-order transition
\begin{eqnarray}\label{jump}
\Delta = \mu_+ = - \frac{3 q_2}{2 q_3} \;,
\end{eqnarray}
which is clearly positive only when $q_2 < 0$ (given that $q_3>0$). }
\item[$(ii)$]{Consider now the case where $q_1<0, \; q_2>0$ and $q_3>0$, which corresponds to the right panel of Fig. \ref{Fig_tri}. Here, when $q_1$ changes sign from negative to positive, i.e., when $q_1=0$, the minimum $\mu_+$ vanishes continuously, signalling a second-order transition.}
\end{itemize}

Finally, a tricritical point in the phase diagram occurs when the first-order transition merges with the second-order transition. 
This happens when the first-order jump discontinuity vanishes, i.e., at $q_2 = 0$. Thus the locus of the tricritical point is then given 
by $q_1 =0$ and $q_2 = 0$.  

\begin{figure}[t]
\includegraphics[width = 0.8\linewidth]{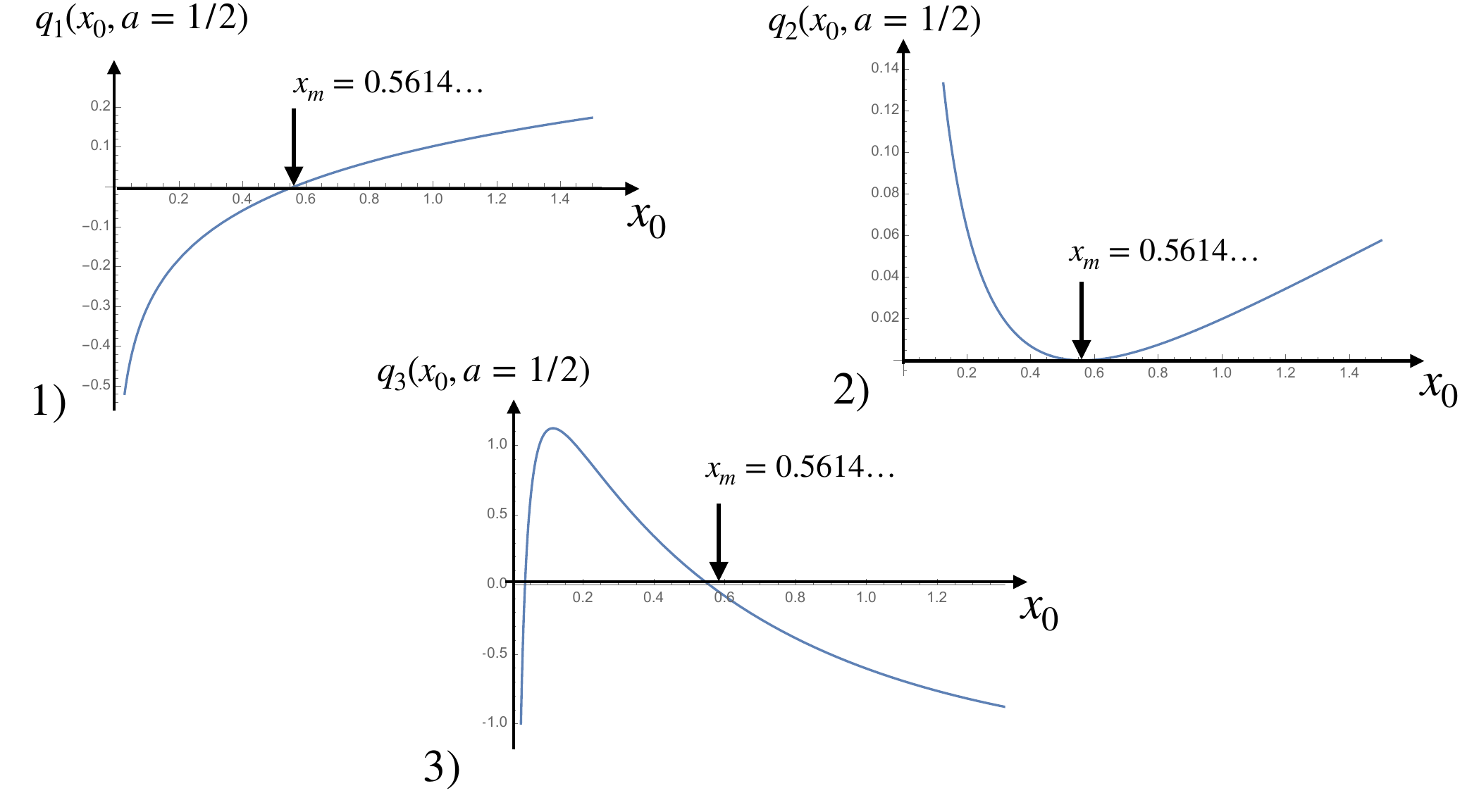}
\caption{The coefficients $q_1(x_0,a)$ in Eq. (\ref{q1}), $q_2(x_0,a)$ in Eq. (\ref{q2}) and $q_3(x_0,a)$ in Eq. (\ref{q3_expl}) plotted as a function of $x_0$ for fixed $a=1/2$. For $x_0=x_m=e^{-\gamma_E} = 0.5614\ldots$, both $q_1$ and $q_2$ vanish while $q_3<0$.}\label{Fig_q123}
\end{figure}
This standard scenario does not hold in our case. We recall that here the underlying parameters are $x_0$ and $a$, and we are investigating how the minimum $\mu^{\star}_{cap}$ of $\widetilde Q_\mu(x_0,a)$ behaves in the $(x_0,a)$ plane (see Fig. 2 of the main text). In the next section, we will carry out the small $\mu$ expansion of $\widetilde Q_\mu(x_0,a)$, starting from the exact Pollaczeck-Spitzer formula and this will allow us to obtain the coefficients $q_i$'s as explicit functions of $x_0$ and $a$. Here we present the general scenario that emerges and allows us to access the tricritical point explicitly as functions of these coefficients $q_i$'s. It turns out from these calculations that, as $x_0$ increases, for fixed $a$, the coefficient $q_1<0$ for $x_0 < x_m=e^{-\gamma_E}$ and is positive for $x_0>x_m$ (see panel 1) of Fig. \ref{Fig_q123}). The coefficient $q_2$, in contrast, is always positive, except at $x_0 = x_m$ where it also vanishes (see panel 2) of Fig. \ref{Fig_q123}). Furthermore, at $x_0=x_m$, the coefficient $q_3$ turns out to be negative for $a$ smaller than some critical value $a_1$ (see panel 3) of Fig. \ref{Fig_q123} where the case $a=1/2$ is represented). Therefore, if we want to investigate the phase diagram in the vicinity of $x_0=x_m$ and $a<a_1$, we need to keep terms up to order $O(\mu^4)$ in the small-$\mu$ expansion (\ref{Qq}). This is because when $q_3<0$, in order to have a nonzero minimum, the function $\widetilde Q_\mu$ must have a $q_4\,\mu^4/4!$ term with $q_4>0$. In this case, assuming $a<a_1$ where $q_3 < 0$ and $x_0 > x_m$ (the last condition ensures that $q_1>0$), there is a nontrivial global minimum at $\mu^{*}$ (see Fig. 3a of the Main Text). As $x_0$ hits $x_c(a)>x_m$, the height of this global minimum becomes equal to $\widetilde Q_{\mu=0}$, so that the minimum at $\mu^*$ coexists with the one at $\mu=0$. When $x_0$ exceeds $x_c(a)$, the minimum jumps to $\mu^*=0$, signalling a first-order phase transition. This jump discontinuity at $x_0 = x_c(a)$ can again be computed from the pair of equations analogous to Eq. (\ref{cond_first}), namely,
\begin{eqnarray}\label{cond_first_4}
\widetilde Q_{\mu=\mu_+} = \widetilde Q_{\mu = 0} \; &\Longrightarrow& \; q_1 \mu + q_2 \frac{\mu^2}{2} + q_3 \frac{\mu^3}{6} + q_4 \frac{\mu^4}{24}= 0 \\
\frac{\partial \widetilde Q_{\mu}}{\partial \mu} \Big \vert_{\mu =0} = 0 \; &\Longrightarrow& \; q_1 + q_2 \mu + q_3 \frac{\mu^2}{2} + q_4 \frac{\mu^3}{6} = 0 \;.
\end{eqnarray} 
Solving this pair of equations, one gets the jump-discontinuity
\begin{eqnarray}\label{jump_4}
\Delta(a) = \left.\frac{2}{3\, q_4} \left( 2\,|q_3| + \sqrt{4\,q_3^2 - 9 \, q_2 q_4}\right)\right|_{x_0=x_c(a)} \;.
\end{eqnarray}
If we now approach the point $(x_0 = x_m, a=a_1)$ in the phase diagram, both coefficients $q_2$ and $q_3$ vanish, indicating that the jump discontinuity $\Delta$ also vanishes. This is indeed the tricritical point. Indeed, for $a\geq a_1$, the critical curve $x_c(a)$ freezes to $x_c(a) = x_m$ (as in Fig. 2 of the Main Text).

\subsection{Computation of the coefficients of the Landau expansion for small $\mu$}

To obtain the coefficients $\{q_i\}$ above, one needs to expand Eq. (\ref{PS2}) in powers of $\mu$ and invert, if possible, the corresponding coefficients back to the real space variable $x_0$. Noting that $\lim_{\mu\rightarrow0} k^{\mu}=1$ for all $k$ (except in a vanishing region near $k=0$ which does not contribute to the integral), one obtains
\begin{equation}\label{q0l}
\left.\int_0^{\infty}\widetilde{Q}_{\mu}(x_0,a)e^{-\lambda x_0} dx_0\right|_{\mu=0}=
\frac{1}{\lambda\sqrt{1-a}}\exp\left[ -\frac{1}{2}
\ln[1-ae^{-1}]\right],
\end{equation}
where we have used the identity $\int_0^{\infty}\frac{1}{\lambda^2+k^2}dk=\frac{\pi}{2\lambda}$. Denoting the inverse Laplace transform as ${\cal L}_{\lambda \to x_0}^{-1}$ and using the fact that ${\cal L}_{\lambda \to x_0}^{-1}(1/\lambda)=1$, one deduces
\begin{equation}\label{q0}
q_0(x_0,a)=\left.\widetilde{Q}_{\mu}(x_0,a)\right|_{\mu=0}=\frac{1}{\sqrt{(1-a)(1-ae^{-1})}},
\end{equation}
as reported in the Main Text. Notably, this expression is independent of the starting position $x_0$.

To calculate $q_1$, we start by taking the first derivative of Eq. (\ref{PS2}) and use  $\partial_{\mu}e^{-k^{\mu}}=-k^{\mu}e^{-k^{\mu}}\ln k$. Exchanging the order of the derivative and integral operators, one obtains
\begin{equation}
\left.\int_0^{\infty}\frac{\partial\widetilde{Q}_{\mu}(x_0,a)}{\partial \mu}\right|_{\mu=0}e^{-\lambda x_0} dx_0=-\frac{ae^{-1}}{\pi\sqrt{1-a}(1-ae^{-1})^{3/2}}\int_0^{\infty}\frac{\ln k}{\lambda^2+k^2}dk = - \frac{ae^{-1}}{2\sqrt{1-a}(1-ae^{-1})^{3/2}} \frac{\ln \lambda}{\lambda} \;.
\end{equation}
Using the inverse Laplace transform 
\begin{equation}\label{LT2}
{\cal L}_{\lambda \to x_0}^{-1}\left(\frac{\ln \lambda}{\lambda}\right)= - (\gamma_E + \ln x_0) \;,
\end{equation}
where $\gamma_E = 0.5772 \ldots$ is the Euler constant. This gives
\begin{equation}\label{q1}
q_1(x_0,a)=\left.\frac{\partial\widetilde{Q}_{\mu}(x_0,a)}{\partial \mu}\right|_{\mu=0}=\frac{ae^{-1}}{2\sqrt{1-a}(1-ae^{-1})^{3/2}}(\gamma_E+\ln x_0),
\end{equation}
which is the expression (7) of the Main Text. A plot of $q_1(x_0,a)$ vs $x_0$ for fixed $a=1/2$ is given in  Fig. \ref{Fig_q123}.
The coefficient $q_1$ changes from negative to positive as $x_0$ crosses the value $x_m=e^{-\gamma_E}$, independently of $a$.

To calculate $q_2$, we derive twice Eq. (\ref{PS2}) with respect to $\mu$, set $\mu=0$, and obtain after some steps
\begin{equation}
\left.\int_0^{\infty}\frac{\partial^2\widetilde{Q}_{\mu}(x_0,a)}{\partial \mu^2}\right|_{\mu=0}e^{-\lambda x_0} dx_0=\frac{a^2\sqrt{e}}{\sqrt{1-a}(e-a)^{5/2}}\left[\frac{\lambda}{\pi^2}
\left(\int_0^{\infty}\frac{\ln^2k}{\lambda^2+k^2}dk
\right)^2+\frac{1}{\pi}\int_0^{\infty}\frac{\ln^2 k}{\lambda^2+k^2} \right].
\end{equation}
We next make use of the identities $\int_0^{\infty}\frac{\ln k}{\lambda^2+k^2}dk=\frac{\pi\ln \lambda}{2\lambda}$ and $\int_0^{\infty} \frac{\ln^2 k}{\lambda^2+k^2}dk=\frac{\pi^3+4\pi\ln^2\lambda}{8\lambda}.$ The above expression becomes
\begin{equation}
\left.\int_0^{\infty}\frac{\partial^2\widetilde{Q}_{\mu}(x_0,a)}{\partial \mu^2}\right|_{\mu=0}e^{-\lambda x_0} dx_0=\frac{a^2\sqrt{e}}{\sqrt{1-a}(e-a)^{5/2}}\left[\frac{3\ln^2\lambda}{4\lambda}+\frac{\pi^2}{8\lambda} \right].
\end{equation}
The r.h.s. of this equation can be inverted exactly owing to the fact that 
\begin{equation} \label{LT_1}
{\cal L}_{\lambda \to x_0}^{-1}\left(\frac{\ln^2\lambda}{\lambda}\right)=(\gamma_E+\ln x_0)^2-\frac{\pi^2}{6} \;.
\end{equation}
One then finds,
\begin{equation}\label{q2}
q_2(x_0,a)=\left.\frac{\partial^2\widetilde{Q}_{\mu}(x_0,a)}{\partial \mu^2}\right|_{\mu=0}=\frac{3a^2\sqrt{e}}{4\sqrt{1-a}(e-a)^{5/2}}(\gamma_E+\ln x_0)^2,
\end{equation}
as in Eq. (8) of the Main Text. A plot of $q_2(x_0,a)$ vs $x_0$ for fixed $a=1/2$ is shown in Fig. \ref{Fig_q123}.  
It turns out that $q_2(x_0,a)$ also vanishes at $x_0=x_m$, for all $a$, as for $q_1(x_0,a)$. We then need to proceed to the third order calculation to conclude about the curvature of $\widetilde{Q}_{\mu}$ at $\mu=0$.

We derive Eq. (\ref{PS2}) three times with respect to $\mu$ and set $\mu=0$. The resulting expression can be written as
\begin{equation}
\int_0^{\infty}\left.\frac{\partial^3 \widetilde{Q}(x_0,a)}{\partial\mu^3} \right|_{\mu=0}e^{-\lambda x_0}dx_0=\frac{1}{\lambda\sqrt{(1-a)(1-ae^{-1})}}(I_3+3I_1I_2+I_1^3),
\end{equation}
where 
\begin{equation}
I_{\ell}=\int_0^{\infty}\left.\frac{\partial^{\ell} h_{\mu}(k)}{\partial \mu^{\ell}}\right|_{\mu=0}dk
\end{equation}
with 
\begin{equation}
h_{\mu}(k)=-\frac{\lambda}{\pi}\frac{\ln(1-ae^{-k^{\mu}})}{\lambda^2+k^2}.
\end{equation}
These integrals can be calculated exactly and the expression simplified with the help of Mathematica. One obtains
\begin{equation} \label{third_deriv}
\int_0^{\infty}\left.\frac{\partial^3 \widetilde{Q}(x_0,a)}{\partial\mu^3} \right|_{\mu=0}e^{-\lambda x_0}dx_0=
-\frac{a\sqrt{e}}{16\sqrt{1-a}(e-a)^{7/2}}\left(G(a)\frac{\ln \lambda}{\lambda} +F(a)\frac{\ln^3 \lambda}{\lambda}\right),
\end{equation}
where
\begin{eqnarray}
&&G(a)=3\pi^2(3a^2+4ea-2e^2 )\label{G_a}\\
&&F(a)=2(11a^2+8ea-4e^2) \label{F_a} \;.
\end{eqnarray}
The inverse Laplace transform of the first term is again ${\cal L}_{\lambda \to x_0}^{-1}\left(\frac{\ln\lambda}{\lambda}\right)=-\gamma_E-\ln x_0$ from Eq. \eqref{LT2}, which vanishes at $x_0=x_m$. To invert the second term, we first define a function $W(x)$ such that
\begin{eqnarray} \label{f_of_x}
\int_0^\infty W(x)\, e^{-\lambda x}\, dx = \frac{\ln^3 \lambda}{\lambda} \;.
\end{eqnarray}
Directly inverting this Laplace transform is difficult. We actually use a convolution trick as follows. First, 
we take a derivative with respect to $\lambda$ to obtain
\begin{eqnarray} \label{derivative}
\int_0^\infty x\, W(x)\, e^{-\lambda x}\, dx  &=& \frac{1}{\lambda^2} \ln^3 \lambda -  \frac{3}{\lambda^2} \ln^2 \lambda \\
&=& \frac{\ln \lambda }{\lambda} \; \frac{\ln^2 \lambda}{\lambda}  - \frac{3}{\lambda} \; \frac{\ln^2 \lambda}{\lambda} \;.
\end{eqnarray}
Expressed in this form, we can now invert each term on the r.h.s. using the convolution theorem and the results from Eqs. \eqref{LT2} and \eqref{LT_1}. This gives
\begin{eqnarray}\label{expl_Fx}
W(x) = - \frac{1}{x} \int_0^x dy\, (\gamma_E + \ln y)\left( \left(\gamma_E + \ln(x-y)\right)^2 - \frac{\pi^2}{6}\right) - \frac{3}{x} \int_0^x dy\, \left( (\gamma_E + \ln y)^2- \frac{\pi^2}{6}\right) \;.
\end{eqnarray}
Performing the integral using Mathematica, we get
\begin{equation} \label{LT3}
{\cal L}_{\lambda \to x_0}^{-1} \left( \frac{\ln^3 \lambda}{\lambda}\right) = -\frac{1}{2} \left( \gamma_E + \ln x_0\right)\left(2 \gamma_E^2 - \pi^3 + 2 \ln x_0 \left(2 \gamma_E + \ln x_0 \right) \right) - 2 \zeta(3) \;,
\end{equation}
where $\zeta(3) = 1.20205690 \ldots$ is the Ap\'ery's constant. Inserting this result in Eq. \eqref{third_deriv} we get
\begin{eqnarray}\label{q3_expl}
q_3(x_0,a)&=& \frac{\partial^3 \tilde Q_\mu(x_0,a)}{\partial \mu^3} \Big \vert_{\mu =0}  \nonumber \\
&=& -\frac{a\sqrt{e}}{16\sqrt{1-a}(e-a)^{7/2}}\left[ \left( \gamma_E + \ln x_0\right) \left( G(a) + \frac{1}{2}F(a)\left(2 \gamma_E^2 - \pi^3 + 2 \ln x_0(2\gamma_E+\ln x_0)\right)\right) + 2 \zeta(3)\right],
\end{eqnarray}
where $G(a)$ and $F(a)$ are given in Eqs. \eqref{G_a} and \eqref{F_a} respectively. A plot of $q_3(x_0,a)$ vs $x_0$ for fixed $a=1/2$ is shown in Fig. \ref{Fig_q123}. Evaluating this at $x_0=x_m = e^{-\gamma_E}$, we get a simpler expression
\begin{equation}\label{q3}
\left. q_3(x_0=x_m,a)=\frac{\partial^3 \widetilde{Q}(x_m,a)}{\partial\mu^3} \right|_{\mu=0}=\frac{a\sqrt{e}K}{8\sqrt{1-a}(e-a)^{7/2}}(11a^2+8ea-4e^2),
\end{equation}
where 
\begin{equation} \label{K}
 K\equiv-\left.{\cal L}_{\lambda \to x_0}^{-1}\left(\frac{\ln^3\lambda}{\lambda}\right)\right|_{x_0=x_m}=2\zeta(3) = 2.4041138 \ldots \;.
\end{equation}
We checked this result numerically using the Gaver-Stehfest method described in Section \ref{appStehfest}. The result in Eq. \eqref{q3} appeared in the Main Text as Eq. (9). The polynomial $F(a)$ has only one root in the interval $[0,1]$, given by
\begin{equation}
a_1=\frac{2e}{11}(\sqrt{15}-2)=0.925690...
\end{equation}
Hence a tricritical point is located at $(a_1,x_m)$ in the $(a,x_0)$-plane. For $0<a<a_1$, $q_3(x_0=x_m,a)<0$ and the transition is first order, while for $a_1<a<1$, $q_3(x_0=x_m,a)>0$ and the transition is second order. In

\section{Expansion of the conditional MFPT $T_{\mu}(x_0,s=a)$ in powers of $\mu$}\label{LandauT}

Recalling the identity that relates the CMFPT to $\widetilde{Q}_{\mu}$,
\begin{equation}\label{Tgen}
T_{\mu}(x_0,a)
=a\frac{\partial}{\partial a}\ln\left[1-(1-a)\widetilde{Q}_{\mu}(x_0,s=a) \right],
\end{equation}
it is possible to expand $T_{\mu}(x_0,a)$ in powers of $\mu$,
\begin{equation}\label{Tt}
T_{\mu}(x_0,a)=t_0(x_0,a) +t_1(x_0,a)\mu+\frac{1}{2!}t_2(x_0,a)\mu^2+\frac{1}{3!}t_3(x_0,a)\mu^3+...
\end{equation}
from the knowledge of
the coefficients $\{q_i\}$ previously determined.
The leading term follows from Eq. (\ref{q0}),
\begin{equation}
t_0(x_0,a)=T_{\mu=0}(x_0,a)=\frac{a(1-e^{-1})}{2\left[\sqrt{(1-ae^{-1})(1-a)}-1+a\right](1-ae^{-1})},
\end{equation}
and is independent of $x_0$.

\begin{figure}[t]
\includegraphics[width = 0.6\linewidth]{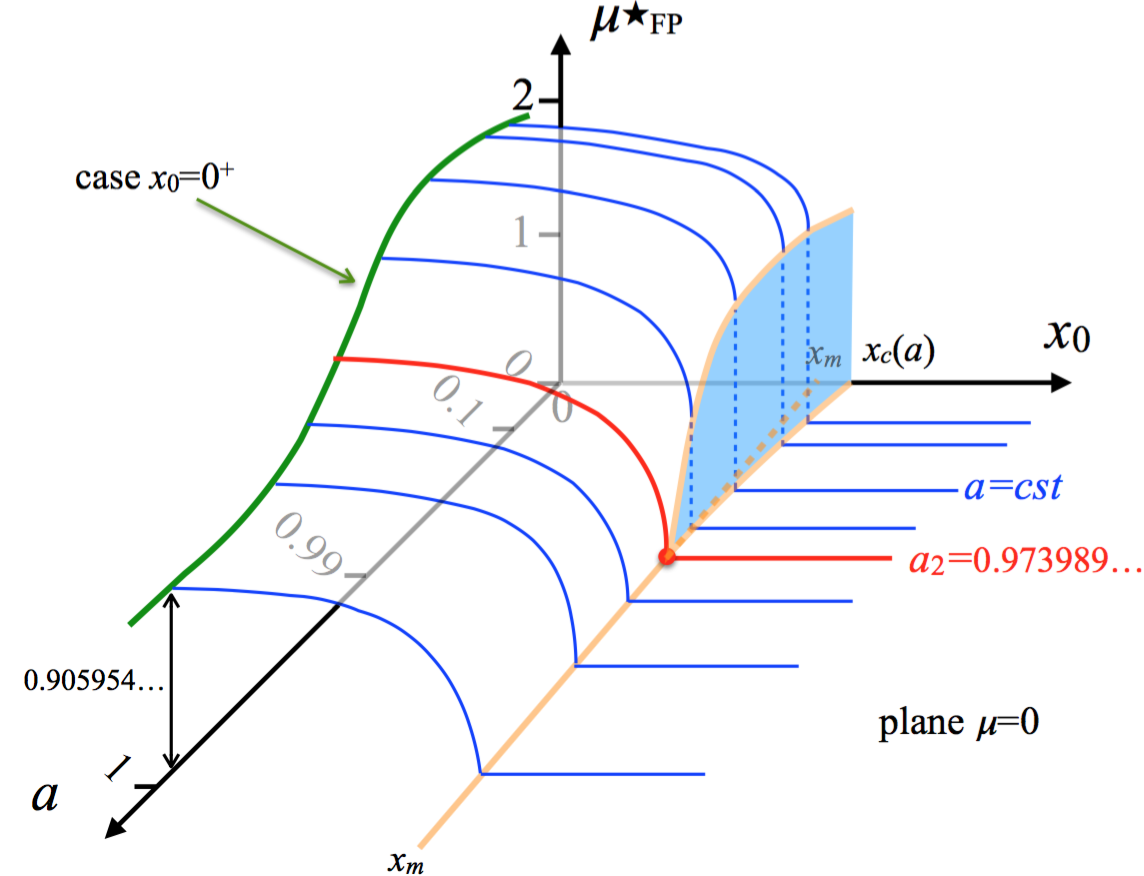}
\caption{Phase diagram for the L\'evy index $\mu^{\star}_{FP}$ that minimizes the conditional mean first passage time, as a function of the living probability $a$ and the initial position $x_0$.}\label{Fig_diag_FP}
\end{figure}

It is easy to see that the two following terms $t_1$ and $t_2$ will have the same behaviors with respect to $x_0$ than $q_1$ and $q_2$, respectively:
\begin{equation}
t_1(x_0,a)=\left.\frac{\partial T_{\mu}(x_0,a)}{\partial \mu}\right|_{\mu=0}\propto \ln x_0+\gamma_E
\end{equation}
 and
\begin{equation}
t_2(x_0,a)=\left.\frac{\partial^2 T_{\mu}(x_0,a)}{\partial \mu^2}\right|_{\mu=0}\propto (\ln x_0+\gamma_E)^2.
\end{equation}
The proportionality constants in the two above expressions are functions of $a$ only.
Therefore the slope at the origin $t_1(x_0,a)$ also changes sign at the position $x_0=x_m$ for all $a$ [like $q_1(x_0,a)$], while $t_2(x_0,a)$ vanishes at the same point, too. We therefore need to proceed to the third order to determine whether $T_{\mu}(x_m,a)$ increases or decreases with $\mu$ at small $\mu$. By taking the third derivative of Eq. (\ref{Tgen}), setting $\mu=0$ and then $x_0=x_m$, only the terms involving $q_0$ and $q_3$ are non-zero,
\begin{equation} 
t_3(x_0=x_m,a)=\frac{\partial^3 T_{\mu}(x_0,a)}{\partial \mu^3}\Big |_{\mu=0,x_0=x_m}=-\frac{a\partial_{a}((1-a)q_3)}{[1-(1-a)q_0]}\Big |_{x_0=x_m}-\frac{a(1-a)q_3\partial_a((1-a)q_0)}{[1-(1-a)q_0]^2}
\Big |_{x_0=x_m}.
\end{equation}
Using the expressions for $q_0$ and $q_3$ given in Eq. (\ref{q0}) and (\ref{q3}), respectively, one obtains,
\begin{eqnarray}\label{t3}
t_3(x_0=x_m,a)&=&
\frac{a\sqrt{e}K}{16(1-(1-a)q_0)^2(e-a)^{5}}\Big\{a(e-1)\sqrt{e}
(11a^2+8ea-4e^2)\nonumber\\
&&+(\sqrt{e}-\sqrt{e-a}/\sqrt{1-a})
\left[(11-93e)a^3+6(15-4e)ea^2
+12e^2(e+1)a-8e^3\right] \Big\} \;,
\end{eqnarray}
where $K = 2 \zeta(3)$ as in Eq. \eqref{K}. Eq. (\ref{t3}) has one unique root in the interval $(0,1)$, which is found numerically to be $a_2=0.973989...$
For $0<a<a_2$, $t_3(x_0=x_m,a)<0$ and the transition is discontinuous, while for $a_2<a<1$,  $t_3(x_0=x_m,a)>0$ and the transition is continuous.
In Figure \ref{Fig_diag_FP}, we summarize the properties of the optimal exponent in the $(x_0,a)$ plane.

\section{Optimal strategy when $a\rightarrow1$}

In the limit of permanent targets, $a\rightarrow1$, the Pollaczek-Spitzer formula (\ref{PS2}) indicates that $\widetilde{Q}_{\mu}(x_0,a)\simeq g_{\mu}(x_0)/\sqrt{1-a}$ where $g_{\mu}(x_0)$ is a function of order one. While $\widetilde{Q}_{\mu}$ diverges as $(1-a)^{-1/2}$,
the capture probability in Eq. (2) of the Main Text takes the form
\begin{equation}\label{Ca1}
C_{\mu}(x_0,a)=\frac{[1-(1-a)\widetilde{Q}_{\mu}(x_0,a)]}{a}\simeq1-\sqrt{1-a}\;g_{\mu}(x_0)+{\cal O}(1-a) \;,
\end{equation}
which is very close to unity independently of the choice of $\mu$. Regarding the CMFPT, the situation is quite different and the choice of $\mu$ is important even in the limit $a\rightarrow 1$, as shown below. From Eq. (\ref{Tgen}), and using $\widetilde{Q}_{\mu}(x_0,a)\simeq g_{\mu}(x_0)/\sqrt{1-a}$, 
we find
\begin{equation}\label{TQa1}
T_{\mu}(x_0,a) \simeq \frac{1}{2\sqrt{1-a}}\;g_{\mu}(x_0).
\end{equation}
The CMFPT is large, although much smaller than the target lifetime $(1-a)^{-1}$. Notice that $g_{\mu}(x_0)$ appears at leading order in the expression of $T_{\mu}$, instead of being a sub-leading contribution as in Eq. (\ref{Ca1}) for $C_{\mu}$. This function has a non-trivial behavior, with significant variations with respect to $\mu$, at fixed $x_0$. Consequently its minimization makes sense for the vast majority of trajectories that capture the target.
The small $\mu$ behavior of $g_{\mu}(x_0)$ up to third order is straightforwardly obtained from Eqs. (\ref{Qq}), (\ref{q0}), (\ref{q1}), (\ref{q2}) and (\ref{q3}), and its minimization yields $\mu_{FP}^{\star}$ (or $\mu_{cap}^{\star}$). Introducing the reduced position to the transition point as
\begin{equation}
\epsilon\equiv\frac{x_0-x_m}{x_m},
\end{equation}
we have, for $|\epsilon|\ll 1$,
\begin{equation}\label{gmu}
g_{\mu}(x_0)=\frac{1}{\sqrt{1-e^{-1}}}+\frac{e^{-1}\epsilon}{2(1-e^{-1})^{3/2}}\mu+\frac{K(11+8e-4e^2)}{8e^3(1-e^{-1})^{7/2}}\frac{\mu^3}{3!}+...,
\end{equation}
where we have neglected the term $q_2$ which is of order $\epsilon^2$.
The minimization of this quantity with respect to $\mu$ gives
\begin{equation}
\mu_{cap}^{\star}=\mu_{FP}^{\star}=0\quad {\rm if}\quad \epsilon\ge0,
\end{equation}
and 
\begin{equation}
\mu_{cap}^{\star}=\mu_{FP}^{\star}=2(e-1)\sqrt{\frac{2}{K(11+8e-4e^2)}}\ |\epsilon|^{1/2} \simeq 1.7549 |\epsilon|^{1/2} \quad {\rm if}\quad \epsilon<0
\end{equation}
Hence there exists a non-trivial optimal search strategy of long-lived targets for $x_0<x_m$. The optimal L\'evy exponent is independent of the target lifetime and persists in the limit of permanent targets. Evaluating the numerical constants appearing in Eq. (\ref{gmu}), one obtains from Eq. (\ref{TQa1}),
\begin{equation}
T_{\mu}(x_0,a)\simeq\frac{1}{\sqrt{1-a}}(0.62888+0.18299\epsilon\mu+0.01980\mu^3+...).
\end{equation}
At the optimum parameter $\mu=\mu_{FP}^{\star}$, the CMFPT is
\begin{equation}
T_{\mu}^{\star}(x_0,a)\simeq\frac{1}{\sqrt{1-a}}(0.62888-0.21411|\epsilon|^{3/2}+...).
\end{equation}
for $\epsilon<0$.

\section{Concavity approximation of the Pollaczeck-Spitzer formula}

We start with the Pollaczeck-Spitzer, given in Eq. (4)-(5) of the Main Text. As mentioned earlier, the main difficulty with this exact formula is that, inverting the Laplace transform with respect to $\lambda$ is very hard. To circumvent this difficulty, we use a concavity approximation. Using the concavity of the logarithm, {\it i.e.}, the inequality  $\sum_i w_i \ln(r_i) \le \ln( \sum_i w_i r_i)$ for any set of positive reals $r_i$ and normalized  positive weights satisfying $\sum_i w_i = 1$, one gets
\begin{equation}\label{log}
\lambda\int_0^{\infty}\ln[\widetilde{Q}(x_0,s)]e^{-\lambda x_0}dx_0\le \ln\left[\lambda\int_0^{\infty} \widetilde{Q}(x_0,s)e^{-\lambda x_0}dx_0\right] \;.
\end{equation}
By replacing the $\le$ above by an ``$=$'' sign, one can perform the inverse Laplace transform in Eq. (4) and deduce a general expression
\begin{equation}\label{concav}
\widetilde{Q}_{\mu,approx}(x_0,s)= \frac{1}{\sqrt{1-s}}e^{-\frac{1}{\pi}\int_0^{\infty}\ln[1-s\hat{f}(k)]\frac{\sin(kx_0)}{k}dk} \;,
\end{equation}
where we have used the identity ${\cal L}^{-1}[k/(\lambda^2+k^2)]=\sin(kx_0)$ for $x_0>0$. Eq. (\ref{concav}) is easy to evaluate numerically. Interestingly, the first two terms of its small $\mu$ expansion coincide with the exact expressions $q_0$ and $q_1$ given respectively in Eqs. (\ref{q0}) and (\ref{q1}), as well as the first terms terms of its small $x_0$ expansion discussed below. Consequently, Eq.~(\ref{concav}) gives the correct slope-change point $x_m$ and also captures qualitatively the different curves in Figs. 3a, 3b and 3c of the Main Text.

\section{Small $x_0$ behavior of $\widetilde{Q}_{\mu}(x_0,a)$ and $\mu_{cap}^{\star}(x_0,a)$}

The case $x_0\ll 1$ can be studied from the large $\lambda$ expansion of $\widetilde{Q}(x_0,s)$ (see \cite{kusmierz2014first} for a similar calculation in the context of resetting processes). 
The fact that the first passage properties of the walk depend on the step length distribution for $x_0>0$ implies that the optimization process becomes non-trivial already for close-by targets. Let us expand Eq. (\ref{PS2}) up to second order in $1/\lambda$,
\begin{equation}\label{PSlargel}
\int_0^{\infty}\widetilde{Q}(x_0,s)e^{-\lambda x_0} dx_0\simeq
\frac{1}{\lambda\sqrt{1-s}} -\frac{1}{\lambda^2\pi\sqrt{1-s}}
\int_0^{\infty}\ln\left(1-se^{-|bk|^{\mu}}\right) dk+\mathcal{O}(1/\lambda^3), 
\end{equation}
which is inverted as
\begin{equation}\label{Qsmallx0}
\widetilde{Q}(x_0,s)\simeq
\frac{1}{\sqrt{1-s}} -x_0\ 
\frac{1}{\pi\sqrt{1-s}}
\int_0^{\infty}\ln\left(1-se^{-|k|^{\mu}}\right) dk+\mathcal{O}(x_0^2). 
\end{equation}
Notice that the same first order expansion is obtained from the concavity approximation [Eq. (12) of Main Text].
Using the general expression for the capture probability [Eq. (2) in Main Text], one deduces from Eq. (\ref{Qsmallx0})
\begin{equation}\label{Csmallx0}
C_{\mu}(x_0,a)\simeq
\frac{1}{1+\sqrt{1-a}} +x_0\ {\cal T}_{\mu}(a) +\mathcal{O}(x_0^2). 
\end{equation}
with
\begin{equation}\label{Tmu}
{\cal T}_{\mu}(a)=\frac{\sqrt{1-a}}{a\pi}
\int_0^{\infty}\ln\left[1-ae^{-|k|^{\mu}}\right] dk.
\end{equation}
One checks that $C_{\mu}(x_0=0^+,a)$ is independent of the jump distribution, which is a consequence of the Sparre Andersen theorem \cite{andersen1954fluctuations}. One also notices that $C_{\mu}(x_0=0^+,a)<1$ (except if $a=1$), since the walker must cross the origin (take a strictly negative position) to find the target, which may disappear before this happens. If the target is alive during only the first step ($a=0$), then $C_{\mu}(x_0=0^+,a=0)=1/2$, as expected from the symmetry of the step length distribution.

\begin{figure}[t]
\centering
			\includegraphics[width=0.42\textwidth]{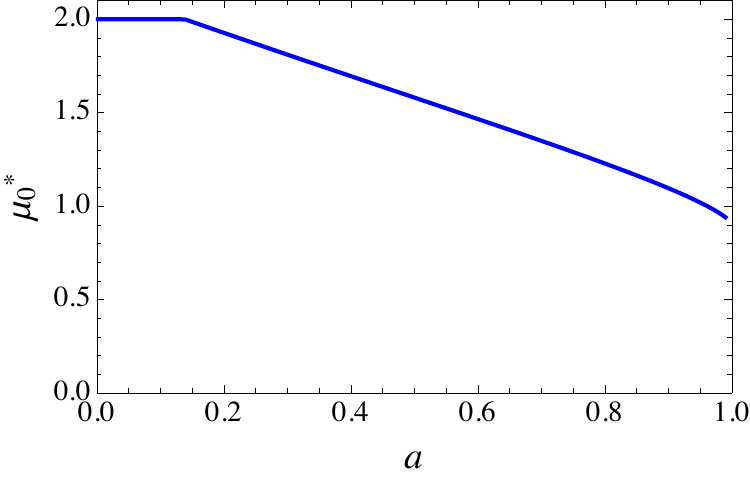}
   \includegraphics[width=0.55\textwidth]{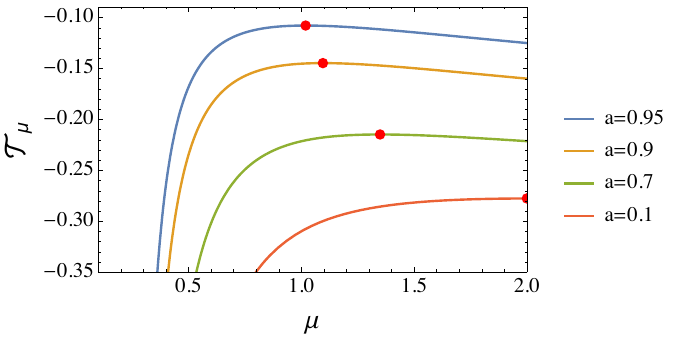}
			\caption{(Right:) Optimal L\'evy exponent as a function of the living probability $a$ for walks starting very close to the origin ($x_0\ll 1$). (Left:) $\mu$-dependent correction to the capture probability in this regime for several $a$ [see Eqs. (\ref{Csmallx0})-(\ref{Tmu})]. The dots indicate the maxima at $\mu_0^*$. }
			\label{fig:x0small}
\end{figure}

For $0<x_0\ll1$, the optimal exponent value is the one that maximizes ${\cal T}_{\mu}(a)$. It is independent of $x_0$ and we re-note $\mu_{cap}^{\star}(x_0\ll 1 ,a)$ as $\mu_0^*(a)$.
Setting $\partial_{\mu}{\cal T}_{\mu}(a)=0$, one obtains a transcendental equation,
\begin{equation}\label{mustar0}
\int_0^{\infty} dk\frac{k^{\mu_0^*} \ln k}{e^{k^{\mu_0^*}}-a}=0.
\end{equation}
By solving numerically Eq. (\ref{mustar0}), one finds that $\mu_0^*(a)$ decreases monotonically,  as displayed in Figure \ref{fig:x0small}-Left. In particular, $\mu_0^*$ tends to a non-trivial value for long-lived targets,
\begin{equation}
\lim_{a\rightarrow 1}\mu_0^*(a)=0.905954...,\quad 
\end{equation}
which is rather close to the Cauchy case $\mu=1$. 
Let us keep in mind that the largest acceptable exponent is $\mu=2$, which corresponds the the Gaussian distribution. Interestingly this value is reached at a particular $a_G>0$. Setting $\mu_0^*=2$ into Eq. (\ref{mustar0}) one finds
\begin{equation}
a_G=0.1381...
\end{equation}
Therefore the optimal distribution remains Gaussian for smaller living probabilities, {\it i.e.}, $\mu_0^*(a)=2$ in the interval $0\le a\le a_G$. We conclude that if the searcher starts very close to the target, Gaussian walks are optimal for very short-lived targets, whereas L\'evy flights are advantageous for long-lived targets.

Of course, in the limit $a\rightarrow1$, the prefactor in front of the integral in Eq. (\ref{Tmu}) vanishes, therefore the advantage brought by the optimal exponent compared to other values of $\mu$ becomes negligible and all the strategies have a $C_{\mu}(x_0,a)$ very close to $1$. The non-monotonic (monotonic) variations of the corrective factor ${\cal T}_{\mu}(a)$ for $0<a<a_0$ ($a>a_0$, respectively) are shown in Figure \ref{fig:x0small}-Right.

%